\documentclass{article}

\usepackage{epsfig} 
\newlength{\abstractwidth} 
\renewcommand{\thefootnote}{\fnsymbol{footnote}} 
\renewcommand{\thanks}[1]{\footnote{#1}} % Use this for footnotes 
\newcommand{\starttext}{ 
\setcounter{footnote}{0} 
\renewcommand{\thefootnote}{\arabic{footnote}}} 
\renewcommand{\theequation}{\thesection.\arabic{equation}} 
\newcommand{\be}{\begin{equation}} 
\newcommand{\bea}{\begin{eqnarray}} 
\newcommand{\eea}{\end{eqnarray}} 
\newcommand{\ee}{\end{equation}} 
 
\newcommand{\Oplus}{$\mathbf{O}\llap+ \: $}

\def\12{{1 \over 2}}

\newcommand{\tab}{\hspace{5mm}}

\begin{document} 
\renewcommand{\theequation}{\thesection.\arabic{equation}} 
\begin{titlepage} 
\bigskip
\centerline{\Large \bf {Group Structure of an Extended Poincare Group}}
\bigskip
\begin{center} 
{\large James Lindesay\footnote{
Permanent address, Department of Physics, Howard University, Washington, DC 20059}
} \\
\end{center}
\bigskip\bigskip 
\begin{abstract} 
In previous papers we extended the Lorentz and Poincare groups to include a set of
\textit{Dirac boosts} that give a direct correspondence with a set of
generators which for spin 1/2 systems are proportional to the Dirac matrices.  The
groups are particularly useful for developing general linear wave equations
beyond spin 1/2 systems.  In this paper we develop explicit group properties
of the extended Poincare group to obtain group parameters that will be
useful for physical calculations in systems which manifest the
group properties.  The inclusion of space-time translations will allow future
explorations of the gauge properties inherent in the group structure.

\end{abstract} 
\end{titlepage} 
%format for equation
%\be 
%S={Area \over 4G} ={4 \pi R^2 \over 4G}. 
%\label{entropy} 
%\ee 
%format for citation 
%\cite{dscft}  
%format for figures
%\threefigures{gr1}{gr2}{gr3}{Large time behavior of correlation functions for (a) N=5, (b) N=10 and (c) N=20}
\starttext \baselineskip=17.63pt \setcounter{footnote}{0} 
%%%%%%%%%%%%%%%% 
\setcounter{equation}{0} 
\section{ Introduction} 
\tab 
One of the most useful properties of Dirac's equation\cite{Dirac} for spinor fields
is that the equations satisfy a linear energy-momentum relationship in the field
equations.
In previous papers\cite{jlspinor}\cite{xPoincare}\cite{xLgroup} we have developed a
set of group generators and field equations that generalize Dirac's formulation. 
The equations have appropriate correspondence with the Dirac equation for
spin 1/2 systems.

In this paper, we will develop the group structure elements for the extended Poincare
group developed in reference\cite{xPoincare}.  We will build on the extended Lorentz
group structure elements developed in reference\cite{xLgroup}. 
We will explicitly calculate those group structure
elements that will be relavant for calculations involving systems which have
the extended Poincare group as a local gauge pre-symmetry.  In future submissions
we will examine the quantum geometrodynamics of systems with
extended Poincare pre-symmetry.

\setcounter{equation}{0} 
\section{Group Theoretic Conventions} 
\tab 
We will continue to utilize the group conventions and parameterization developed
for the extended Lorentz subgroup of this extended Poincare group developed in
reference \cite{xLgroup}. 
In particular, the generators of infinitesimal transformations for operator
representations are given by
\be
\left . i \, \mathbf{X}_r \: \equiv \: 
{\partial \over \partial \mathcal{M}^r} \, S(\mathcal{M}) \right |_{\mathcal{M}=\mathcal{I}} .
\label{generators}
\ee
and Lie structure matrices are defined by
\be
\mathbf{\Theta}_r ^s (\mathcal{M}) \: \equiv  \: 
\left . {\partial \Phi^s (\mathcal{M'};\mathcal{M})  \over 
\partial \mathcal{M'}^r } \right |_{\mathcal{M'}=\mathcal{I}},
\label{LieStructure}
\ee
where $\Phi^s (\mathcal{M'};\mathcal{M})$ is the group composition (closure)
element resulting from the operation $\mathcal{M'}$ on element $\mathcal{M}$.
The generators transform under the representations of the group
as given by the relation
\be
S(\mathcal{M}^{-1}) \, \mathbf{X}_r \, S(\mathcal{M}) \: = \:
{\mathbf{O}\llap+}_r ^s (\mathcal{M}) \: \mathbf{X}_s
\label{Opluscalculate}
\ee
where the matrices \Oplus given by
\be
{\mathbf{O}\llap+}_r ^s (\mathcal{M}) \: \equiv \:
\left . {\partial \over \partial \mathcal{M'}^r}
\Phi^s (\mathcal{M}^{-1} \, ; \, \Phi(\mathcal{M'} \, ; \, \mathcal{M}))
\right | _{\mathcal{M'}=\mathcal{I}}
\ee
form a fundamental representation of the group. 
The Lie structure matrices are useful to determine transformation
properties of gauge fields\cite{Mickens}.

We will therefore calculate the group matrix elements \Oplus  and  $\mathbf{\Theta}$ 
that determine the transformation properties of generators and gauge
fields for this group.  In subsequent papers, these elements will be used
in developing dynamical models for physical systems.

\setcounter{equation}{0} 
\section{Extended Group Commutation Relations} 
\tab
As developed previously, the commutation relationships between the generators of this extended
Lorentz group are given by
\bea
\left [ J_j \, , \, J_k \right] \: = \: i \, \epsilon_{j k m} \, J_m 
\label{JJeqn} \\
\left [ J_j \, , \, K_k \right] \: = \: i \, \epsilon_{j k m} \, K_m 
\label{JKeqn} \\
\left [ K_j \, , \, K_k \right] \: = \: -i \, \epsilon_{j k m} \, J_m 
\label{KKeqn} \\
\left [ \Gamma^0 \, , \, \Gamma^k \right] \: = \: i \, K_k 
\label{Gam0Gameqn} \\
\left [ \Gamma^0 \, , \, J_k \right] \: = \: 0 
\label{Gam0Jeqn} \\
\left [ \Gamma^0 \, , \, K_k \right] \: = \: -i \,  \Gamma^k 
\label{Gam0Keqn} \\
\left [ \Gamma^j \, , \, \Gamma^k \right] \: = \: -i \, \epsilon_{j k m} \, J_m 
\label{GamGameqn} \\
\left [ \Gamma^j \, , \, J_k \right] \: = \: i \, \epsilon_{j k m} \, \Gamma^m 
\label{GamJeqn} \\
\left [ \Gamma^j \, , \, K_k \right] \: = \: -i \, \delta_{j k} \, \Gamma^0 
\label{GamKeqn}
\eea
An extended Lorentz group Casimir operator can be constructed in the form
\be
C_\Lambda \: = \: \underline{J} \cdot \underline{J} \,-\, \underline{K} \cdot \underline{K}
\,+\, \Gamma^0 \, \Gamma^0 \,-\, \underline{\Gamma} \cdot \underline{\Gamma}.
\label{Casimir_Lam}
\ee

For the extended Poincare group,
the non-vanishing extended translation commutators involving the operators
$\hat{P}_\mu$ and $\hat{\mathcal{G}}$ are given by
\bea
\left [ J_j \, , \, P_k \right] \: = \: i \, \epsilon_{j k m} \, P_m 
\label{JPeqn} \\
\left [ K_j \, , \, P_0 \right] \: = \: -i \, P_j 
\label{KP0eqn} \\
\left [ K_j \, , \, P_k \right] \: = \: -i \, \delta_{j k} \, P_0 
\label{KPeqn} \\
\left [ \Gamma^\mu \, , \, P_\nu \right] \: = \: \pm i \, \delta_\nu ^\mu \, \mathcal{G} 
\label{GamPeqn} \\
\left [ \Gamma^\mu \, , \, \mathcal{G} \right] \: = \: \pm i \, \eta^{\mu \nu} \, P_\nu 
\label{GamGeqn}
\eea
where the upper signs were used in the previous results\cite{xPoincare}.
We can construct an extended Poincare group Casimir operator given by
\be
\mathcal{C}_{(\mu)} \: \equiv \: \mathcal{G}^2 \,-\, \eta^{\beta \nu} P_\beta P_\nu .
\label{Casimir_mu}
\ee

To be consistent with the group element $\mathcal{G}$ being the generator
for representations as defined in Equation \ref{generators},
we will choose the lower signs in Equation \ref{GamGeqn}
\bea
\left [ \Gamma^\mu \, , \, P_\nu \right] \: = \: - i \, \delta_\nu ^\mu \, \mathcal{G} 
\label{GamPeqn2} \\
\left [ \Gamma^\mu \, , \, \mathcal{G} \right] \: = \: - i \, \eta^{\mu \nu} \, P_\nu 
\label{GamGeqn2}
\eea

\setcounter{equation}{0} 
\section{Extended Poincare Group Structure} 
\tab 
We will construct finite element transformations in the extended Poincare group
in what follows.  This group will have rotations, general Lorentz transformations,
the extended Lorentz transformations, and extended translations as subgroups. 
A general group element is characterized by the 15 parameters given by
$\{ \alpha, \vec{a}, \vec{\omega}, \underline{u}, \underline{\theta}  \}$, where
$\alpha$ is the element conjugate to the extended scalar translation generator $\mathcal{G}$,
$a^\mu$ is the element conjugate to the momentum $P_\mu$,
$\omega_\mu$ is the element conjugate to the Dirac boost generator $\Gamma^\mu$,
$\beta^m$ (which is directly related to the four-velocity parameter $u^m$) 
is the element conjugate to the Lorentz boost generator $K_m$, and
$\theta^m$ is the element conjugate to the angular momentum $J_m$.

\subsection{General Lorentz Transformations} 
\tab 
The Lorentz group is a subgroup of the extended Poincare group being developed. 
The four-velocity $\vec{u}$ is as usual defined using $u^0 = \sqrt{1+|\underline{u}|^2}$. 
The three-components of the four-velocity are related to the parameters conjugate to
the (Lorentz) boost generators $K_m$ by $\underline{u} \equiv \hat{\beta} \: tanh \beta$.
We will establish our convention for the
Lorentz transformation matrices on four-vectors. The general Lorentz transformation
matrix is defined by 
\be
\Lambda^\mu _{ \quad \nu}(\underline{u}, \underline{\theta}) \equiv 
\mathcal{L}_{\Gamma^\mu} ^{\quad \Gamma^\beta}(\underline{u}) \: 
\mathcal{R}_{\Gamma^\beta} ^{\quad \Gamma^\nu}(\underline{\theta}),
\ee
where 
$\mathcal{L}_{\Gamma^\mu} ^{\quad \Gamma^\beta}(\underline{u}) \equiv 
\mathcal{L}_{\mu} ^{\beta}(\underline{u}) $ and
$\mathcal{R}_{\Gamma^\beta} ^{\quad \Gamma^\nu}(\underline{\theta}) \equiv 
\mathcal{R}_{\beta} ^{\nu}(\underline{\theta})$
are directly related to the 4-Lorentz boost and rotation matrices defined
in reference\cite{xLgroup}.  Since we now have 4-generators with both
covariant and contravariant properties, we will carefully exhibit 
all transformation properties.
This matrix has the usual properties of a Lorentz transformation, leaving invariant
the group metric $\eta_{_{\Gamma^\mu \Gamma^\nu}}$ obtained
from the structure constants of the extended
Lorentz group, which is proportional to the usual Minkowski metric $\eta_{\mu \nu}$. 
The inverse of this matrix can directly be shown to result from lowering and
raising its indeces using the Minkowski metric.
The inverse element of this representation of the Lorentz group satisfies
\be
\{ \underline{u}, \underline{\theta} \}^{-1} \: = \:
 \{ -R(-\underline{\theta})\underline{u}, -\underline{\theta}   \} \: = \:
 \{ -\underline{u} R^{-1}(\underline{\theta}^{-1}), -\underline{\theta}   \}
\ee
The form of the infinitesimal 4-Lorentz generators
\be
\begin{array}{c}
\left ( \mathcal{J}_m \right ) ^\mu _{\quad \nu} \, \equiv \,
\left . {\partial \over \partial \theta_m} \Lambda^\mu _{ \quad \nu}(\underline{0}, \underline{\theta})
\right |_{\underline{\theta}=\underline{0}} \\ \\
\left ( \mathcal{K}_m \right ) ^\mu _{\quad \nu} \, \equiv \,
\left . {\partial \over \partial u_m} \Lambda^\mu _{ \quad \nu}(\underline{u}, \underline{0})
\right |_{\underline{u}=\underline{0}}
\end{array}
\ee
has non-vanishing elements given by
\be
\begin{array}{c}
\left ( \mathcal{J}_m \right ) ^j _{\quad k}  \: = \: \epsilon_{mjk} \\ \\
\left ( \mathcal{K}_m \right ) ^0 _{\quad k} \: = \: 
- \delta_{m,k} \: = \: \left ( \mathcal{K}_m \right ) ^k _{\quad 0} .
\end{array}
\ee

\subsection{Extended Translations} 
\tab 
Group transformations parameterized by elements $a^\mu$
conjugate to the generators $P_\mu$  and $\alpha$ conjugate to the
generator $\mathcal{G}$ will be referred to as
extended translations.  The extended translations all mutually
commute, so that this subset is an abelian subgroup.

\subsubsection{Extended Lorentz Transformations on Extended Translations}
\tab
We examine the effect of extended Lorentz transformations on the 
extended translations. 
Utilizing Equation \ref{Opluscalculate} for this representation, we can read
off various elements of the fundamental representation.  For pure Lorentz
tranformations, we obtain new matrix elements in addition to those obtained
previously for the extended Lorentz group:
\be
\begin{array}{l}
{\mathbf{O}\llap+}_{P_\mu}^{P_\beta} (0,\vec{0},\vec{0},\underline{u},\underline{\theta}) \: = \: 
{\mathbf{O}\llap+}_{\Gamma^\beta}^{(XL)\Gamma^\mu} 
(\vec{0},\underline{u}^{-1},\underline{\theta}^{-1}) \: = \:
\Lambda^\beta _{\quad \mu}(\underline{u}^{-1},\underline{\theta}^{-1}) \: = \:
\Lambda_\beta ^{\quad \mu}(\underline{u},\underline{\theta})  \\ \\
{\mathbf{O}\llap+}_{\mathcal{G}}^{\mathcal{G}} 
(0,\vec{0},\vec{0},\underline{u},\underline{\theta}) \: = \: 1
\end{array}
\label{LorentzonXT}
\ee
where ${\mathbf{O}\llap+}_{\Gamma^\beta}^{(XL)\Gamma^\mu} 
(\vec{\omega},\underline{u},\underline{\theta})$ are the fundamental representation
matrices for the extended Lorentz group as given in reference \cite{xLgroup}.  These
relations demonstrate that $\Gamma^\mu P_\mu$ and $\mathcal{G}$ are Lorentz invariants.

For pure Dirac boosts, the additional fundamental representation elements are given by
\be
\begin{array}{l}
{\mathbf{O}\llap+}_{P_\mu}^{P_\beta} (0,\vec{0},\vec{\omega},\underline{0},\underline{0}) \: = \: 
\delta_\mu ^\beta +(cos(\omega)-1) { \omega_\mu \omega^\beta
\over \omega_\nu \omega^\nu} \, \equiv \, \mathcal{S}_\mu ^\beta (-\vec{\omega})\\ \\
{\mathbf{O}\llap+}_{\mathcal{G}}^{P_\beta} (0,\vec{0},\vec{\omega},\underline{0},\underline{0}) \: = \: 
-sin(\omega) {\omega^\beta \over \omega} 
 \, \equiv \, \mathcal{S}_\mathcal{G} ^\beta (-\vec{\omega})\\ \\
{\mathbf{O}\llap+}_{P_\mu}^{\mathcal{G}} (0,\vec{0},\vec{\omega},\underline{0},\underline{0}) \: = \: 
-sin(\omega) {\omega_\mu \over \omega}
 \, \equiv \, \mathcal{S}_\mu ^\mathcal{G} (-\vec{\omega}) \\ \\
{\mathbf{O}\llap+}_{\mathcal{G}}^{\mathcal{G}} (0,\vec{0},\vec{\omega},\underline{0},\underline{0}) \: = \: 
cos(\omega)  \, \equiv \, \mathcal{S}_\mathcal{G} ^\mathcal{G} (-\vec{\omega}) .
\end{array}
\label{DiraconXT}
\ee
where $\omega^\beta \equiv \eta^{\beta \nu} \omega_\nu$ and
$\omega \:\equiv \: \sqrt{-\vec{\omega} \cdot \vec{\omega}} \: = \:
\sqrt{-\omega_\nu \omega^\nu}$.

\subsubsection{Extended Translations on Extended Lorentz Transformations}
\tab
There are additional fundamental representation matrix elements due to
the non-vanishing commutation of the extended translation generators with
the Dirac boost generators.  Most of these elements are unity on the diagonal. 
The only other non-vanishing elements are given by
\be
\begin{array}{l}
{\mathbf{O}\llap+}_{\Gamma^\mu}^{P_\beta} (\alpha,\vec{0},\vec{0},\underline{0},\underline{0}) \: = \: 
\alpha \eta^{\mu \beta} \\ \\
{\mathbf{O}\llap+}_{\Gamma^\mu}^{\mathcal{G}} 
(0,\vec{a},\vec{0},\underline{0},\underline{0}) \: = \: a^\mu
\end{array}
\ee

The general fundamental transformation matrix for the extended
Poincare group is then given by
\be
\begin{array}{c}
{\mathbf{O}\llap+}_r ^s (\alpha,\vec{a},\vec{\omega},\underline{u},\underline{\theta}) \: = \: 
{\mathbf{O}\llap+}_r ^q (\alpha,\vec{0},\vec{0},\underline{0},\underline{0}) \,
{\mathbf{O}\llap+}_q ^p (0,\vec{a},\vec{0},\underline{0},\underline{0}) \,
{\mathbf{O}\llap+}_p ^n (0,\vec{0},\vec{\omega},\underline{0},\underline{0}) \,
{\mathbf{O}\llap+}_n ^m (0,\vec{0},\vec{0},\underline{u},\underline{0}) \,
{\mathbf{O}\llap+}_m ^s (0,\vec{0},\vec{0},\underline{0},\underline{\theta}) \\
\quad \quad \: = \: {\mathbf{O}\llap+}_r ^q (\alpha,\vec{0},\vec{0},\underline{0},\underline{0}) \,
{\mathbf{O}\llap+}_q ^p (0,\vec{a},\vec{0},\underline{0},\underline{0}) \,
{\mathbf{O}\llap+}_p ^{(XL) \, s} (\vec{\omega},\underline{u},\underline{\theta})
\end{array}.
\ee
where ${\mathbf{O}\llap+}_p ^{(XL) \, s}$ are extended Lorentz group representation
matrices from reference\cite{xLgroup} expanded to include Equations
\ref{LorentzonXT} and \ref{DiraconXT}.

\subsection{Extended Poincare Group Transformations} 
\tab 
To develop a description of the group structure of this extended
Poincare group, we will explicitly demonstrate the group composition
rule and Lie structure elements of the complete group.  The representation
we have developed has been constructed by sequential pure rotation, Lorentz boost,
Dirac boost, space-time translation, and extended scalar translation: 
\be
\mathbf{M} (\alpha,\vec{a},\vec{\omega}, \underline{u}, \underline{\theta}) \: \equiv \:
\mathbf{C}(\alpha)\mathbf{V}(\vec{a})
\mathbf{W}(\vec{\omega}) \, \mathbf{L}(\underline{u}) \, \mathbf{R} (\underline{\theta})
\ee
The group composition rule then defines elements in this same manner
\be
\begin{array}{l}
\mathbf{M} (\alpha_2,\vec{a}_2,\vec{\omega}_2, \underline{u}_2, \underline{\theta}_2) \,
\mathbf{M} (\alpha_1,\vec{a}_1,\vec{\omega}_1, \underline{u}_1, \underline{\theta}_1) \: \equiv \: \\
\mathbf{C}(\alpha(\alpha_2, \vec{a}_2, \vec{\omega}_2, \underline{u}_2, \underline{\theta}_2 \, ; \,
\alpha_1, \vec{a}_1,\vec{\omega}_1, \underline{u}_1, \underline{\theta}_1))
\mathbf{V}(\vec{a}(\alpha_2, \vec{a}_2, \vec{\omega}_2, \underline{u}_2, \underline{\theta}_2 \, ; \,
\alpha_1, \vec{a}_1,\vec{\omega}_1, \underline{u}_1, \underline{\theta}_1)) \\
\mathbf{W}(\vec{\omega}(\alpha_2, \vec{a}_2, \vec{\omega}_2, \underline{u}_2, \underline{\theta}_2 \, ; \,
\alpha_1, \vec{a}_1,\vec{\omega}_1, \underline{u}_1, \underline{\theta}_1) ) \, 
\mathbf{L}(\underline{u}(\alpha_2, \vec{a}_2, \vec{\omega}_2, \underline{u}_2, \underline{\theta}_2 \, ; \,
\alpha_1, \vec{a}_1,\vec{\omega}_1, \underline{u}_1, \underline{\theta}_1)) \\
\mathbf{R} (\underline{\theta}(\alpha_2, \vec{a}_2, \vec{\omega}_2, \underline{u}_2, \underline{\theta}_2 \, ; \,
\alpha_1, \vec{a}_1,\vec{\omega}_1, \underline{u}_1, \underline{\theta}_1))
\end{array}
\ee
The group composition elements can be expressed using previously
constructed functions in terms of pure extended Lorentz transformations
and extended translations:
\be
\begin{array}{l}
\alpha (\alpha_2, \vec{a}_2, \vec{\omega}_2, \underline{u}_2, \underline{\theta}_2 \, ; \,
\alpha_1, \vec{a}_1,\vec{\omega}_1, \underline{u}_1, \underline{\theta}_1) \: = \:
\alpha_2 + \alpha_1 \, \left( {\mathbf{O}\llap+}(-\vec{\omega}_2) \right)_\mathcal{G} ^\mathcal{G} +
a_1 ^\nu \left( {\mathbf{O}\llap+}(-\underline{\theta}_2) {\mathbf{O}\llap+}(-\underline{u}_2) 
{\mathbf{O}\llap+} (-\vec{\omega}_2)
 \right)_{P_\nu} ^\mathcal{G} \\ \\
a^\mu (\alpha_2, \vec{a}_2, \vec{\omega}_2, \underline{u}_2, \underline{\theta}_2 \, ; \,
\alpha_1, \vec{a}_1,\vec{\omega}_1, \underline{u}_1, \underline{\theta}_1) \: = \:
a_2 ^\mu + \alpha_1 \, \left( {\mathbf{O}\llap+}(-\vec{\omega}_2) \right)_\mathcal{G} ^{P_\mu} +
a_1 ^\nu \left( {\mathbf{O}\llap+}(-\underline{\theta}_2) {\mathbf{O}\llap+}(-\underline{u}_2) 
{\mathbf{O}\llap+} (-\vec{\omega}_2)
 \right)_{P_\nu} ^{P_\mu} \\ \\
\vec{\omega}(\alpha_2, \vec{a}_2, \vec{\omega}_2, \underline{u}_2, \underline{\theta}_2 \, ; \,
\alpha_1, \vec{a}_1,\vec{\omega}_1, \underline{u}_1, \underline{\theta}_1) \: = \: 
\vec{\omega}^{(XL)} ( \vec{\omega}_2, \underline{u}_2, \underline{\theta}_2 \, ; \,
\vec{\omega}_1, \underline{u}_1, \underline{\theta}_1) \\ \\
\underline{u}(\alpha_2, \vec{a}_2, \vec{\omega}_2, \underline{u}_2, \underline{\theta}_2 \, ; \,
\alpha_1, \vec{a}_1,\vec{\omega}_1, \underline{u}_1, \underline{\theta}_1) \: = \:  
\underline{u}^{(XL)} ( \vec{\omega}_2, \underline{u}_2, \underline{\theta}_2 \, ; \,
\vec{\omega}_1, \underline{u}_1, \underline{\theta}_1) \\ \\
\underline{\theta}(\alpha_2, \vec{a}_2, \vec{\omega}_2, \underline{u}_2, \underline{\theta}_2 \, ; \,
\alpha_1, \vec{a}_1,\vec{\omega}_1, \underline{u}_1, \underline{\theta}_1) \: = \:
\underline{\theta}^{(XL)} ( \vec{\omega}_2, \underline{u}_2, \underline{\theta}_2 \, ; \,
\vec{\omega}_1, \underline{u}_1, \underline{\theta}_1)
\end{array}
\label{EPGcomposition}
\ee
where fundamental representation matrices with only one argument presume the vanishing
of all other elements, and $\vec{\omega}^{(XL)}$, $\underline{u}^{(XL)}$,
and $\underline{\theta}^{(XL)}$ are the extended Lorentz group
composition rules obtained in reference\cite{xLgroup}.
The inverse element can then be shown to be given by
\be
\begin{array}{l}
\{\alpha, \vec{a}, \vec{\omega}, \underline{u}, \underline{\theta} \}^{-1} \: = \:
 \{-\alpha \mathcal{S}_{\mathcal{G}}^{\mathcal{G}} (\vec{\omega}^{-1})
- \vec{a} \cdot  \Lambda^{-1}(\underline{u}^{-1}, \underline{\theta}^{-1}) \cdot
\vec{\mathcal{S}}^\mathcal{G}(\omega^{-1}) , \\
\quad \quad -\alpha \vec{\mathcal{S}}_{\mathcal{G}} (\vec{\omega}^{-1})
- \vec{a} \cdot  \Lambda^{-1}(\underline{u}^{-1}, \underline{\theta}^{-1}) \,
\mathcal{S}(\vec{\omega}^{-1}) , \:
 - \vec{\omega} \, \Lambda(\underline{u}^{-1} , \underline{\theta}^{-1})  , \:
-\underline{u} \, R^{-1}(\underline{\theta}^{-1}), \underline{\theta}^{-1}=-\underline{\theta}   \}
\end{array}
\ee
where $(\vec{a} \, \Lambda^{-1})^\mu=a^\nu \, \Lambda_\nu ^{\quad \mu}$ and
$(\vec{\omega} \, \Lambda)_\mu=\omega_\nu \, \Lambda^\nu _{\quad \mu}$.
More explicitly, the inverse extended translation elements can be expressed
\be
\begin{array}{l}
\alpha^{-1} \:= \: -\alpha \left( {\mathbf{O}\llap+} 
(-\vec{\omega}^{-1})  \right) _\mathcal{G} ^ \mathcal{G}
- a^\nu \left( {\mathbf{O}\llap+} (-\underline{\theta}^{-1}) {\mathbf{O}\llap+} (-\underline{u}^{-1})
{\mathbf{O}\llap+} (-\vec{\omega}^{-1}) \right) _{P_\nu} ^\mathcal{G} \\ \\
(a^{-1})^\mu \:= \: -\alpha \left( {\mathbf{O}\llap+} 
(-\vec{\omega}^{-1})  \right) _\mathcal{G} ^ {P_\mu}
- a^\nu \left( {\mathbf{O}\llap+} (-\underline{\theta}^{-1}) {\mathbf{O}\llap+} (-\underline{u}^{-1})
{\mathbf{O}\llap+} (-\vec{\omega}^{-1}) \right) _{P_\nu} ^{P_\mu}
\end{array}
\ee
where representation matrices of single arguments assume all other
arguments vanish. 

\subsection{Lie Structure Elements} 
\tab 
Finally, we can use the composition rules given in Equation \ref{EPGcomposition},
along with the definition given in Equation \ref{LieStructure}, to explicitly develop
the Lie structure matrices
\be
\begin{array}{l}
\mathbf{\Theta}_{\alpha} ^{\alpha} (\alpha, \vec{a}, \vec{\omega},
\underline{u}, \underline{\theta}) \: = \: 1 \\ \\
\mathbf{\Theta}_{\omega_\mu} ^{\alpha} (\alpha, \vec{a}, \vec{\omega},
\underline{u}, \underline{\theta}) \: = \: a^\mu \\ \\
\mathbf{\Theta}_{a^\beta} ^{a^\mu} (\alpha, \vec{a}, \vec{\omega},
\underline{u}, \underline{\theta}) \: = \: \delta_\beta ^\mu \\ \\
\mathbf{\Theta}_{\omega_\beta} ^{a^\mu} (\alpha, \vec{a}, \vec{\omega},
\underline{u}, \underline{\theta}) \: = \: \alpha \, \eta^{\mu \beta} \\ \\
\mathbf{\Theta}_{u^j} ^{a^\mu} (\alpha, \vec{a}, \vec{\omega},
\underline{u}, \underline{\theta}) \: = \: 
\delta_0 ^\mu \, a^j + \delta_k ^\mu \, a^0 \\ \\
\mathbf{\Theta}_{\theta^j} ^{a^k} (\alpha, \vec{a}, \vec{\omega},
\underline{u}, \underline{\theta}) \: = \: \epsilon_{jmk} \, a^m \\ \\
\mathbf{\Theta}_{s} ^{\omega_\mu} (\alpha, \vec{a}, \vec{\omega},
\underline{u}, \underline{\theta}) \: = \: 
\mathbf{\Theta}_{s} ^{(XL) \omega_\mu} (\vec{\omega},
\underline{u}, \underline{\theta}) \\ \\
\mathbf{\Theta}_{s} ^{u^j} (\alpha, \vec{a}, \vec{\omega},
\underline{u}, \underline{\theta}) \: = \: 
\mathbf{\Theta}_{s} ^{(XL) u^j} (\vec{\omega},
\underline{u}, \underline{\theta}) \\ \\
\mathbf{\Theta}_{s} ^{\theta^j} (\alpha, \vec{a}, \vec{\omega},
\underline{u}, \underline{\theta}) \: = \: 
\mathbf{\Theta}_{s} ^{(XL) \theta^j} (\vec{\omega},
\underline{u}, \underline{\theta}),
\end{array}
\ee
where $\mathbf{\Theta} ^{(XL)}$ are the extended Lorentz group
Lie structure matrices given in reference\cite{xLgroup}.
We will be able to utilize these matrices in general
gauge transformations for systems which have a local
gauge symmetry in this group.

\section{Conclusions}
\tab
We have demonstrated a group representation of
the extended Poincare group developed in reference\cite{xPoincare}. 
The fundamental representation matrices and Lie structure matrices
have been explicitly calculated.  The group structure elements
allow direct construction of finite dimensional unitary vector representations under
which quantum states transform as developed in reference\cite{xPoincare}. 
The set of extended translations are invariant under rotations (for massive
systems), which is the little group for appropriately chosen standard states. 
Several Lorentz invariants can be constructed from the set of generators
of this group, which provides a rich unified structure for the dynamics
of physical systems.  The implications of this group structure on the
quantum geometrodynamics of systems exhibiting this symmetry or
pre-symmetry will be further explored in future explorations.

\section{Acknowledgements}
\tab
The author wishes to acknowledge the support of Elnora Herod and
Penelope Brown during the intermediate periods prior to and after
his Peace Corps service (1984-1988), during which time the bulk of this work was
accomplished.  In addition, the author wishes to recognize the
hospitality of the Department of Physics at the University of Dar
Es Salaam during the three years from 1985-1987 in which a substantial portion of
this work was done.


\begin{thebibliography}{999} 
\baselineskip=17pt 
\itemsep = 2pt 
%format for bibliography
%\bibitem{stretch} L. Susskind, L. Thorlacius, J. Uglum, 
%The Stretched Horizon and Black Hole Complementarity, 
%hep-th/9306069, Phys. Rev. D48 (1993) 3743-3761 
%\newline
%C.R. Stephens, G. 't Hooft, B.F. Whiting, 
%Black Hole Evaporation without Information Loss, 
%gr-qc/9310006, Class.Quant.Grav. 11 (1994) 621-648 

\bibitem{Dirac} P.A.M. Dirac,
Proc. Roy. Soc. (London), A117, 610 (1928);
ibid, A118, 351 (1928) 
\bibitem{jlspinor} J. Lindesay,
Linear Spinor Field Equations for Arbitrary Spins,
math-ph/0308003 (2003)
\bibitem{xPoincare} J. Lindesay,
An Extended Poincare Algebra for Linear Spinor Field Equations,
math-ph/0308015 (2003)
\bibitem{xLgroup} J. Lindesay,
Group Structure of an Extended Lorentz Group,
math-ph/0309060 (2003)
\bibitem{Mickens} J.V. Lindesay and H.L. Morrison,
The Geometry of Quantum Flow, in \textit{Mathematical Analysis of Physical Systems},
R.E. Mickens, ed., Van Nostrand Reinhold, New York, p 135 (1985) 
\newline
\end{thebibliography}
\end{document}